# 6G Empowering Future Robotics: A Vision for Next-Generation Autonomous Systems

Mona Ghassemian, Andrés Meseguer Valenzuela, Ana Garcia Armada, Dejan Vukobratovic, Periklis Chatzimisios, Kaspar Althoefer, RR Venkatesha Prasad

*Abstract*— **The convergence of robotics and next-generation communication is a critical driver of technological advancement. As the world transitions from 5G to 6G, the foundational capabilities of wireless networks are evolving to support increasingly complex and autonomous systems. We examine the transformative impact of 6G on enhancing key robotics functionalities. It provides a systematic mapping of IMT-2030 key performance indicators to robotic functional blocks, including sensing, perception, cognition, actuation, and self-learning. Building upon this mapping, we propose a high-level architectural framework integrating robotic, intelligent, and network service planes, underscoring the need for a holistic approach. As an example, use case, we present a real-time, dynamic safety framework enabled by IMT-2030 capabilities for safe and efficient human-robot collaboration in shared spaces.**

## I. INTRODUCTION

The field of robotics is undergoing a major transformation, moving from traditional, isolated systems to intelligent, networked agents capable of complex collaboration and interaction. This shift is closely tied to advances in communication technologies, especially the emergence of 6G. This paper explores how 6G capabilities can empower autonomous agents, focusing on sensing and perception, cognition, actuation, and self-learning.

Foundational research at this early stage of 6G definition and standardization is essential to establish the technological and regulatory groundwork for future robotic systems [1], [2]. New and enhanced IMT-2030 functionalities, including intelligence, context awareness, and hyper-reliable low-latency communication (HRLLC), will support robot–human coexistence, collaborative tasks, and safe navigation. These capabilities are crucial for macro-level operations, such as maneuvering and collision avoidance, as well as micro-level tasks, such as grasping and manipulation.

As 6G evolves toward an AI-native network paradigm, its most transformative impact on robotics lies in distributing intelligence across robots, edge, and cloud infrastructure. AI-native 6G capabilities — edge computing, semantic communication, federated learning, and agentic AI — closely match robotic requirements, boosting collaboration, scalability, and operational efficiency. This positions 6G as a key enabler for emerging domains such as soft, swarm, and intelligent actuation systems, where compute-intensive perception, planning, and learning are offloaded to edge and network services. As a result, robotic platforms become more cost-efficient while still leveraging powerful shared models, coordination mechanisms, and agentic AI hosted in intelligent and network service planes. In healthcare, soft robots can exploit low-latency edge inference for safe, context-aware surgical assistance, while logistics robots can use semantic, task-oriented communication and connected perception to navigate complex environments with high reliability.

The current work provides a holistic approach to 6G-enabled robotics and makes three main contributions. In particular, it (i) proposes the systematic mapping of IMT-2030 features to the four core robotic functionalities (sensing, cognition, actuation, and self-learning), (ii) introduces a multi-plane architecture with a novel data governance layer to address trust and ethical concerns, and (iii) uses dynamic safety zones (DSZs) to validate the proposed architecture for real-time adaptive human-robot interaction. Together, these contributions establish a unique framework that complements ongoing research and standardization efforts with concrete architectural insights.

Mona Ghassemian (corresponding author) is with King's College London, Andrés Meseguer Valenzuela is with ITI, Ana Garcia Armada is with UC3M, Dejan Vukobratovic is with UNS, Periklis Chatzimisios is with International Hellenic University and University of New Mexico, Kaspar Althoefer is with QMUL, and Ranga Rao Venkatesha Prasad is with TUDelft.

## II. 6G TECHNOLOGY AND MAPPING WITH ROBOTICS CORE FUNCTIONALITIES

The International Telecommunication Union (ITU) is driving this vision, and its framework [7] by defining the new enhanced capabilities and usage scenarios for IMT-2030 (the official term for 6G). This framework expands IMT-2020 key performance indicators (KPIs) by newer ones to address emerging use cases.

### A. *Key* 6G KPI categories and targets

The following KPIs outline the critical targets for 6G technology, which is pivotal for advancing robotic capabilities.

**(1) Peak data rate:** Perception and swarm coordination demand the transmission of massive sensory and video streams. Peak data rates of 50–200 Gbps and an average of 300–500 Mbps enable real-time multimodal perception, SLAM fusion, and sustain dense robot fleets [7]. **(2) Latency and Reliability**: Actuation and execution in robotics depend on instantaneous response. Latency targets of 0.1–1 ms support haptic teleoperation, collision avoidance, and corrective actions in dynamic safety zones. Reliability with error rates as low as $10^{-5}$–$10^{-7}$ ensures trustworthy operation in a mission-critical context. **(3) Sensing and Positioning**: Perception and safe navigation require precise environmental awareness.

Positioning accuracy of 1–10 cm and integrated sensing and communications (ISAC)-based mapping provide robots with 360° situational awareness, extending the range of local sensors. **(4) Spectrum Efficiency**: Efficient spectrum use sustains cognition and coordination across fleets. With improvements of 1.5 to 3 times, 6G enables simultaneous robot communication, supports distributed planning and multi-agent reasoning, and ensures seamless cooperation in smart factories, warehouses, and large-scale robotic ecosystems. **(5) AI-related Capabilities**: Self-learning and cognition in robotics depend on AI-native networks. Inference latency and the time to federated learning updates ensure real-time adaptation and cooperative model sharing. With network-embedded AI, robots gain scalable autonomy by offloading planning and decision-making computation. **(6) Sustainability and Inclusiveness**: The long-term deployment of robotic systems requires sustainable and inclusive infrastructure. Reducing carbon footprint per transmitted bit lowers maintenance costs for vertical-specific networks, while global coverage supports robotics in agriculture, disaster response, and healthcare.

These KPIs mark a clear shift from 5G, where robots primarily used wireless connectivity for actuation and, to a lesser extent, perception via connected cameras. In contrast, 6G goes beyond communication by integrating **sensing, localization**, and **AI-native functions** into the network, offering a unified framework for perception, cognition, and actuation. This enables real-time situational awareness, adaptive learning, and cooperative intelligence across distributed robots, leading to autonomous, context-aware, networked robotics. Fig. 1 compares the evolution from 5G to 6G [7], showing 6G features supporting key robotic functions, using Autonomous Mobile Robots (AMRs) as an example. It also reflects the transition toward a distributed multi-plane architecture, where sensing, cognition, planning, and actuation are dynamically coordinated across robotic, intelligent, and network service layers.

*B. Key Technology Enablers for Robotics Functionalities*

The core functionalities for any robotic system are sensing and perception, cognition (reasoning and planning), actuation (and execution), and self-learning [1]. Integrating these blocks is essential to developing autonomous intelligent robotic systems.

**Sensing and Perception in Robotic Systems:** Perception is how a robot observes and interprets its environment from sensory data. It relies on a multimodal world model that supports tasks such as scene segmentation, object detection, activity recognition, and motion prediction. This model encodes geometry, semantics, dynamics, and uncertainty, enabling the robot to localize, reconstruct the scene, and reason about objects, interactions, and future changes. Using active inference, the robot predicts the next state and uses sensory data mainly to correct those predictions, reducing data flow. **6G ISAC**, integrated **edge AI**, immersive communication, and **HRLLC** will be key to fusing multimodal data and improving active inference, enabling a richer shared environmental model.

**Cognition, Reasoning, and Planning:** Complementing perception, these functions provide the intelligence for action by reasoning about the environment and organizing data for better decision-making. Cognitive control systems model the robot and its interactions to improve performance and efficiency. Planning within this framework can run on the robot, at the network edge, or in the cloud. With 6G, planning becomes more dynamic and distributed, especially with agentic AI, enabling complex use cases such as general-purpose humanoid assistants, flexible manufacturing, and large-scale fleet management through seamless cross-layer communication. Semantic communication [4]-[5] will further improve context-aware exchange, reducing the data needed for decision-making.

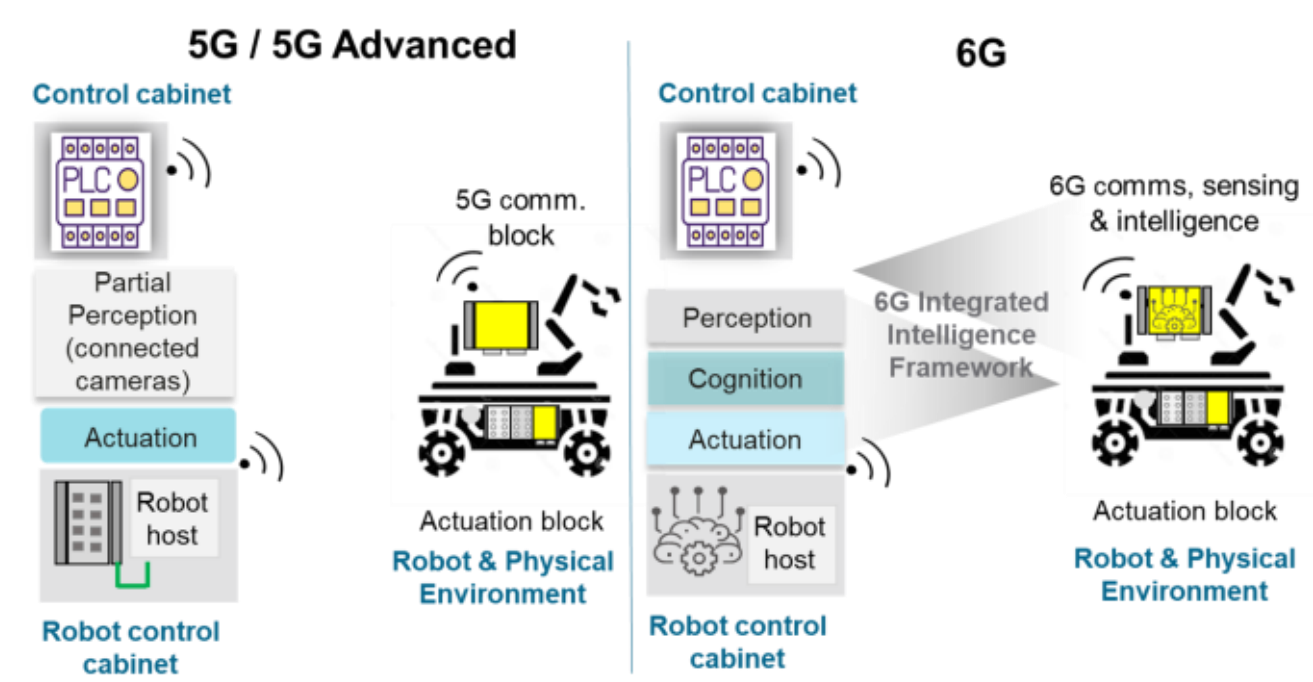

***Figure 1.*** *Evolution from 5G/5GA to 6G for AMRs*

**Actuation and Execution:** This block covers the physical execution of control algorithms using perceptual and cognitive inputs. Effective actuation depends on tight integration between sensing and cognition, with predictive (active) inference playing a key role. By acting predictively rather than purely reactively, systems better adapt to sudden environmental changes. In real time (e.g., a robot adjusting its path to avoid an unexpected obstacle), 6G can minimize delay between sensing and corrective action.

**Self-Learning:** Unsupervised learning helps robots detect patterns and improve their algorithms from experience. Over time, self-learning refines the world model, enhancing predictive performance via active inference. This autonomy enables adaptation through perceptual learning, so robots can improve skills in dynamic, unpredictable settings. With 6G's large data capacity and computing resources, self-learning will accelerate, making robots more adaptable across a wider range of applications. Table I summarizes the key 6G enabling technologies for each robotic functional block.

TABLE I. KEY TECHNOLOGY ENABLERS FOR ROBOTICS FUNCTIONAL BLOCKS AND IMPACT ON FUTURE ROBOTICS

| Functional Blocks | 6G Enablers | Metrics | Robotic Impact |
|---|---|---|---|
| **Sensing and Perception** | ISAC, edge AI, HRLLC | 50–200 Gbps; 300–500 Mbps; 1–10 cm | Shared perception, SLAM, situational awareness |
| **Cognition, Reasoning and Planning** | AI-native, semantic comms, edge/cloud | 1.5–3×spectrum efficiency; AI inference latency | Distributed reasoning & fleet coordination |

| **Execution, and Actuation** | HRLLC, TSN, edge control | 0.1–1 ms; $10^{-5}$–$10^{-7}$ reliability | Teleoperation, collision avoidance, safety |
|---|---|---|---|
| **Self-learning** | FL, edge intelligenc, AI agents | FL update time; energy/bit reduction | Cooperative learning & adaptive autonomy |

## III. Architecture towards Task-Oriented Communications

Traditional networks focus on reliable bit transmission, maximizing throughput, and minimizing latency. For autonomous robot networks, however, the key metric is not the number of bits sent but meaningful, task-relevant information that helps a robot complete a task. Semantic-aware communication addresses this by prioritizing task-relevant information and embedding meaning into the communication process. In the near term, semantic encoding/decoding can appear as an optional processing layer augmenting the packet-based stack, while longer-term evolution may enable tight integration within softwarized L1/L2 functions (e.g., RLC/MAC/PHY) jointly optimized with modulation, coding, and scheduling. This requires redesigning physical and L1/L2 processing and resource management as AI-native semantic-communication blocks.

Existing mechanisms such as scheduling, HARQ, and link adaptation are preserved, but operate on semantically compressed representations and can be extended with task-aware metrics (e.g., semantic fidelity or task success probability) rather than relying solely on bit-level reliability. In practice, this can be realized via configurable data pipes, where semantic processing modules are dynamically instantiated alongside conventional bit pipes in the user plane. For example, perception data from a camera or LiDAR can be converted into task-relevant semantic features like object labels and positions, sent over a semantic-aware data pipe, and reconstructed at the receiver to support control and decision-making.

To fully exploit 6G for robotics, a new architecture is needed. Beyond a simple hardware/software split, it should adopt a multi-plane structure integrating robotic, intelligent, and network services. This aligns with the AI-core concept in ETSI ENI ISG [6], which proposes multi-agent, AI-native control of next-generation network slicing, with agents handling intent processing, slice customization, and closed-loop optimization of 6G resources [6]. Fig. 2 shows a high-level layered 6G framework interconnecting planes composed of the following:

### A. Robotic Vertical Plane

This plane serves as the integration layer for robotic applications. It facilitates services for enhanced perception, cognition, collaborative actions, and the management of dynamic safety zones. It enables the easy adaptation and scaling of robotic systems to various use cases and industries by providing a structured interface for third-party services.

### B. Intelligent Service Plane

This plane, the brain of the system, supports real-time processing of complex environmental data, which is essential for advanced decision-making and operational efficiency. Integrating cognitive communication and computing yields a more intelligent, responsive robotic service plane, where AI is an integral part of the network's functionality rather than just a tool. Self-learning is realized through distributed AI agents at the network and edge that continuously optimize task-oriented slices, QoS profiles, and semantic encoders for robotic workloads [6].

### C. Data Governance Plane

The development and deployment of intelligent robotic systems generate vast amounts of data. The Data Governance Plane manages data collection, storage, and processing, ensuring integrity, accessibility, and reliable provisioning of models and AI tools. It is also critical for ethical, trustworthy, and secure operation, ensuring that data-driven decisions remain transparent and accountable.

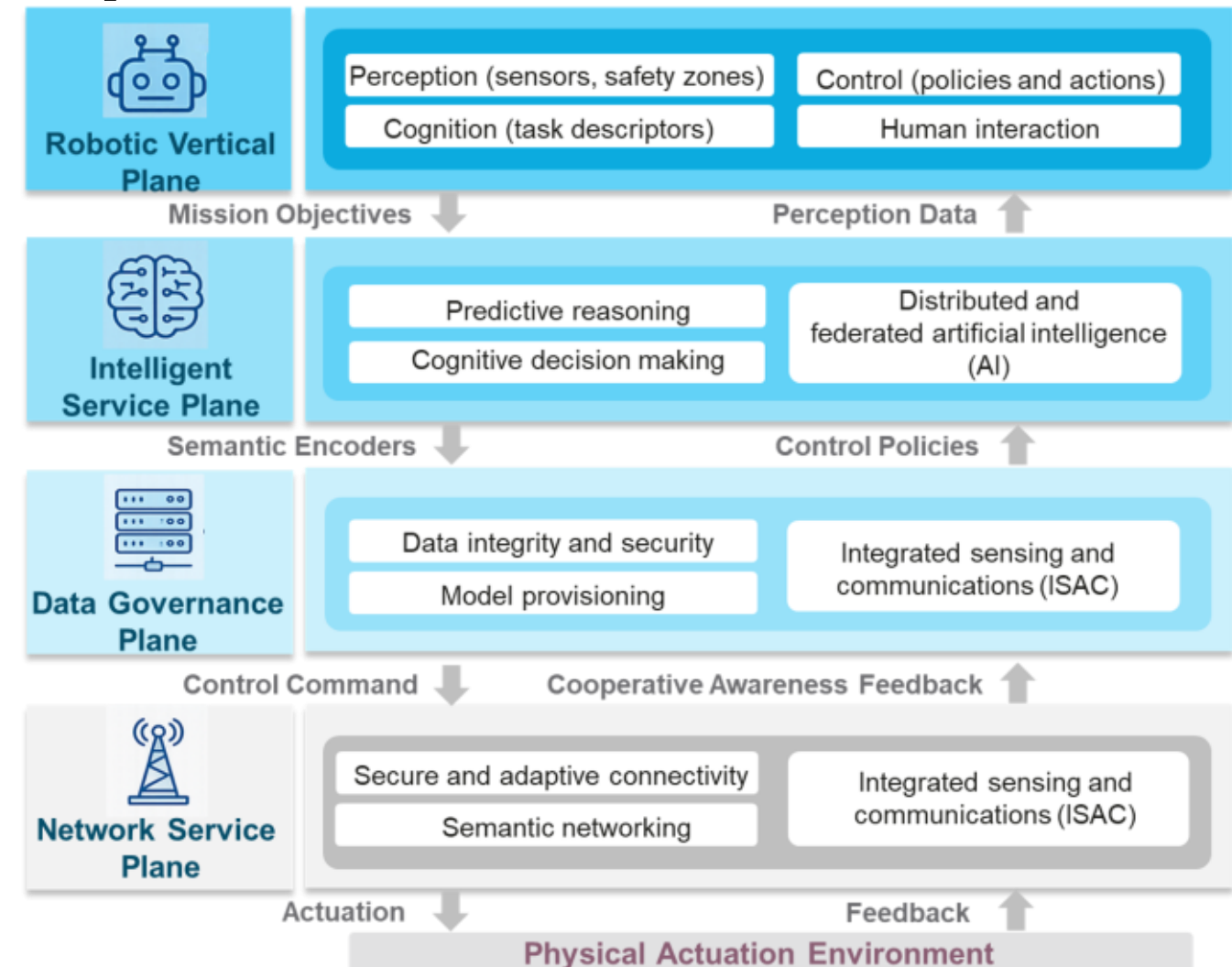

***Figure 2.*** *High-Level Architecture*

### D. Network Service Plane

The network plane is the foundational layer that provides critical sensing and communication capabilities. It encompasses the 3GPP user and control planes, along with a dedicated data plane for exchanging sensing data, distinct from the user plane. AI and ISAC are core features of this plane, enabling the network to communicate with robots and perceive the environment. This plane also includes semantic data representation to improve efficiency and enhance collaboration.

At the interface between the **Robotic Vertical Plane** and the **Intelligent Service Plane**, robotic applications provide task descriptors, semantic intent, and safety constraints (e.g., human–robot separation margins, braking delay limits, or critical perception zones) and receive optimized semantic encoders, control policies, and updated safety zones. This bidirectional exchange supports human–machine interaction by sharing mission-critical needs and returning models and policies refined through perceptual and experiential learning.

The interface between the **Robotic Vertical Plane** and **the**

**Data Governance Plane** mediates access to curated datasets, model artifacts, and feedback for self-learning, while enforcing privacy, trust, and regulatory compliance. Agents request data and submit updated models or logs via governed APIs, keeping continual learning auditable as models evolve.

Finally, the interface between the **Robotic Vertical Plane** and the **Network Service Plane** orchestrates sensing, positioning, computing, and communications as task-oriented services. AI agents can configure and refine slices, semantic L1/L2 pipelines, and ISAC to keep sensing, localization, compute placement, and connectivity aligned with task objectives, using closed-loop feedback to preserve reliability and safety.

These functions are supported by network service handlers (e.g., sensing fusion, communication, and computation offload). Embedding AI agents in the network intelligence plane adds adaptive reasoning and self-learning, while cross-plane security ensures trusted operation. Overall, 6G can unify robotic control and network intelligence to enable real-time perception, cognition, and actuation for next-generation autonomous systems. Table II, below, provides an overview of the functionalities, requirements, and interfaces of the high-level 6G architecture depicted in Fig. 2.

TABLE II. 6G High-Level Architecture Components

| Plane/Function | Handler | Requirements | Example APIs |
| --- | --- | --- | --- |
| **Robotic vertical/ Perception** | Environment perception setup and update | Data collection, real-time updates | Network Sensing APIs; Positioning APIs via NEF/SEAL/CAPIF |
| **Robotic vertical/ Actuation** | Task coordination & execution | Resource coordination, task management | Robotics vertical, human–robot event interfaces |
| **Robotic vertical/ Cognition** | Trajectory & motion planning | Dynamic planning, safety compliance | Mission/schedule APIs; Network-aware planning constraints |
| **Network service plane/ Communication** | Reliable connectivity | UE registration, mobility, slice selection | Core/exposure, AMF / SMF, low-latency slice and NAS APIs |
| **Network service plane/ Sensing** | Multimodal sensing & fusion | Sensor integration, data fusion | RAN/ISAC/UAV nodes APIs; Radio sensing reports |
| **Network service plane/ Computing** | Task computation offload | Resource optimization; Efficient computation | Offload APIs for centralized processing |
| **Network intelligence plane/ AI agents** | Context-aware reasoning & self-Learning | Contextual awareness, learning | AI model hosting, collision-risk APIs |
| **Cross plane/Security** | Authentication and identity management | Secure access, trusted data handling | Security protocols for robotic operations |

## IV. Dynamic Human-Robot Collaboration

To concretize the proposed multi-plane architecture and its interfaces, this section instantiates them in a safety-critical use case: dynamic safety zones for human–robot collaboration. In this use case, 6G capabilities, such as HRLLC, ISAC, and networked sensing, are explicitly mapped to safety functions, communication interfaces, and control loops across the robotic, intelligent, and network service planes.

The safe coexistence of autonomous systems with humans in shared workspaces remains a key barrier to large-scale deployment of mobile robots and collaborative robots (cobots). Static safeguards based on physical fences and hard speed limits are effective but overly conservative, constraining throughput and flexibility in domains such as intralogistics, manufacturing, and healthcare. Collaborative operation, therefore, requires safety mechanisms that are dynamic, perception-driven, and context-aware rather than purely static and rule-based.

Current regulations, such as the EU Machinery Directive and ISO standards [14], impose speed limits and physical safeguards for human-robot collaboration. Although vital for safety, these rules can hinder productivity in speed-critical fields, such as logistics. Enhancing robots' perception beyond on-board sensors offers a solution.

A Dynamic Safety Zone (DSZ) is an adaptive virtual safety region surrounding a robot, human operator, or critical workspace that continuously changes according to contextual information such as robot velocity, trajectory, task criticality, environmental conditions, and human proximity. Unlike traditional static safety zones based on fixed distances or predefined exclusion areas, DSZs are continuously updated in real-time using networked sensing, positioning, and AI-driven prediction mechanisms.

The deployment of DSZs must also comply with regulatory and ethical frameworks. The EU AI Act [14] , ISO 13482[15], and emerging standards emphasize the importance of risk management, transparency, and accountability. Embedding DSZs through compliance-by-design ensures safety conformity and builds user trust, essential in sensitive areas such as healthcare and elder care.

### A. Enabling Dynamic *Safety with 6G ISAC*

Implementing DSZs effectively requires continuous external perception beyond the limitations of local robotic sensors. 6G ISAC capabilities can support real-time monitoring of spatial separation, motion dynamics, and environmental context, enabling adaptive safety enforcement through network-assisted situational awareness [3].

The 6G Radio Access Network (RAN) can sense all objects within predefined zones, with the precision and frequency of measurements adapting to the criticality of the zone, the robot's speed, and human movements. The service request from the robot controller can trigger specific sensing actions via the 6G network to improve measurement accuracy and ensure compliance with safety standards. While ISAC enables this human-robot collaboration, there is a trade-off between the communication and sensing capabilities offered by the radio interface. For instance, models in [15] show trade-offs in coverage, resolution, and hardware requirements.

### B. Architectural *Impact of Dynamic Safety Zones*

In the proposed architecture, DSZ enforcement is distributed across robotic, intelligent, and network service planes, combining local control, AI-driven prediction, and HRLLC/ISAC-enabled communication for safety-critical coordination [3].

To support certifiability, safety-relevant interfaces between planes are defined as explicit safety channels with clear timing, integrity, and availability requirements. For each DSZ flow, the network service plane exposes these channels via standardized northbound APIs that specify: (i) maximum one-way latency, (ii) target packet-delivery reliability, (iii) minimum update frequency, and (iv) degradation policies (e.g., bandwidth reallocation, priority boost, or graceful shutdown) when targets cannot be met. Treating DSZ flows explicitly in this way allows safety requirements from standards to be mapped to concrete 6G configuration parameters, which can then be validated through simulation or analysis.

### C. Safety in Mobile Robotics

Ensuring safety in mobile robotics poses challenges distinct from manipulators, mainly because robots move through large, dynamic, and unstructured spaces. Static measures such as speed limits or exclusion zones are overly restrictive, limiting efficiency and scalability in logistics, healthcare, and service robotics.

DSZs address this by adapting to robot motion and the behavior of nearby entities. For mobile platforms [8], continuously recalculating speed and separation thresholds not only along straight paths but also during turns, stops, and replanning is needed. A Robot Controller can request sensing actions from the 6G Network Infrastructure, highlighting the network's active role in safety protocols.

### *D.* Impact of robotic tasks on QoS Constraints and 6G capabilities

The qualitative safety model above generalizes to different robotic tasks by mapping each task to latency, reliability, sensing accuracy, and throughput constraints, and then dimensioning s6G mechanisms accordingly:

**DSZ-based obstacle avoidance for mobile robots:** In warehouse or hospital-like indoor environments, DSZ updates can follow a 30 Hz sensing rate, as in edge-enabled digital-twin collision-avoidance systems where camera frames are processed at 30 FPS and feature packets are sent each cycle. Closed-loop end-to-end delays of about 68–70 ms in good channels and $\approx$76 ms under degraded Wi-Fi, with average packet error rates around 1% and 0% above 47 dB SNR, correspond to $\gtrsim$99–100% successful DSZ updates [8].

**Collaborative manipulation with DSZ-based speed and separation monitoring**: Isochronous industrial control loops require guaranteed deadlines below 2 ms, while cyclic control traffic tolerates 2–20 ms latency [10]. Mapping cobot SSM to this taxonomy, the DSZ loop is dimensioned as an isochronous flow with reaction times in the 1–2 ms range and very low jitter, carried over TSN class-C Ethernet with sub-millisecond cycle times. This justifies hosting DSZ inference close to the robot controller and reserving dedicated TSN/URLLC resources for safety-critical cobot motions.

**Remote surgery and high-precision teleoperation**: Technical standards recommend a maximum end-to-end delay of 10 ms for haptic-enabled telesurgery, with network experiments over 150–250 km achieving round-trip delays of 4–10 ms on guaranteed links [10]. Overall end-to-end latencies of about 90–95 ms remain clinically usable, though further delay degrades performance. In a 6G-DSZ framework, virtual fixtures and no-go zones would thus be enforced by haptic control loops dimensioned to stay within the 10 ms target whenever possible.

These examples illustrate how the proposed 6G-DSZ architecture can be instantiated with concrete, task-dependent QoS and sensing parameters. While full quantitative validation requires detailed simulation or experiments, the qualitative safety model and explicit task-to-QoS mapping together provide a technically grounded, practically feasible basis for DSZ-enabled 6G systems.

## V. Layered Architectural Framework by example

The proposed four-plane 6G system (Fig. 2) modularly integrates sensing and perception, cognition and planning, and actuation and execution across the robotic, intelligent, data-governance, and network-service planes, and can be instantiated in applications from industrial automation to remote healthcare. Fig. 3 shows how safe human–robot collaboration is realized through these layers, highlighting the exchange of sensor, perception, command, and feedback data between the robots, edge components, and the cloud.

**Interfaces:** The architectural planes define a set of interfaces that connect robotic applications, network intelligence, and data infrastructure, enabling end-to-end self-learning and task-oriented communication.

**Sensing & Perception Layer:** This foundational layer captures the environment and robot state through multimodal sensors (e.g., LiDAR, cameras, microphones, tactile and force sensors, IMUs) and networked sensing/ISAC modules (top layers in Fig. 3). Local or edge processing nodes perform multimodal data fusion, compression, and semantic abstraction so that downstream components operate on compact task-relevant representations rather than raw high-rate streams (for example, a structured description of object poses instead of full-resolution video).

**Cognition, reasoning, and planning layer:** This corresponds to the "Communication", "Planning", and "Reasoning & Self-Learning" blocks (Fig. 3) and acts as the intelligence and coordination hub of the architecture. It orchestrates context-aware data exchange among robots, edge servers, and the cloud using scalable HRLLC, semantic communication mechanisms, and distributed planning algorithms running on digital twins. Within this layer, an intelligent user plane and semantic communication engine jointly exploit AI models to select and adapt traffic, perform task-oriented encoding/decoding, and coordinate multi-robot cooperation, while generative models support predictive control, environment reconstruction, and self-learning from feedback data.

**Control and Execution Layer*:*** The bottom layer in Fig. 3 groups the "Interface & Collaboration", "Actuation", and "Execution & Control" functions and translates high-level decisions into physical motion and human-robot interaction. It contains local controllers, drives, and end-effectors that rely on closed-loop HRLLC connectivity for deterministic execution of critical commands such as grasping, cooperative manipulation, and collision avoidance. Feedback data from tactile, proprioceptive, and HMI interfaces are returned to the cognition layer to refine models and update plans, thereby closing the loop between perception, reasoning, and actuation.

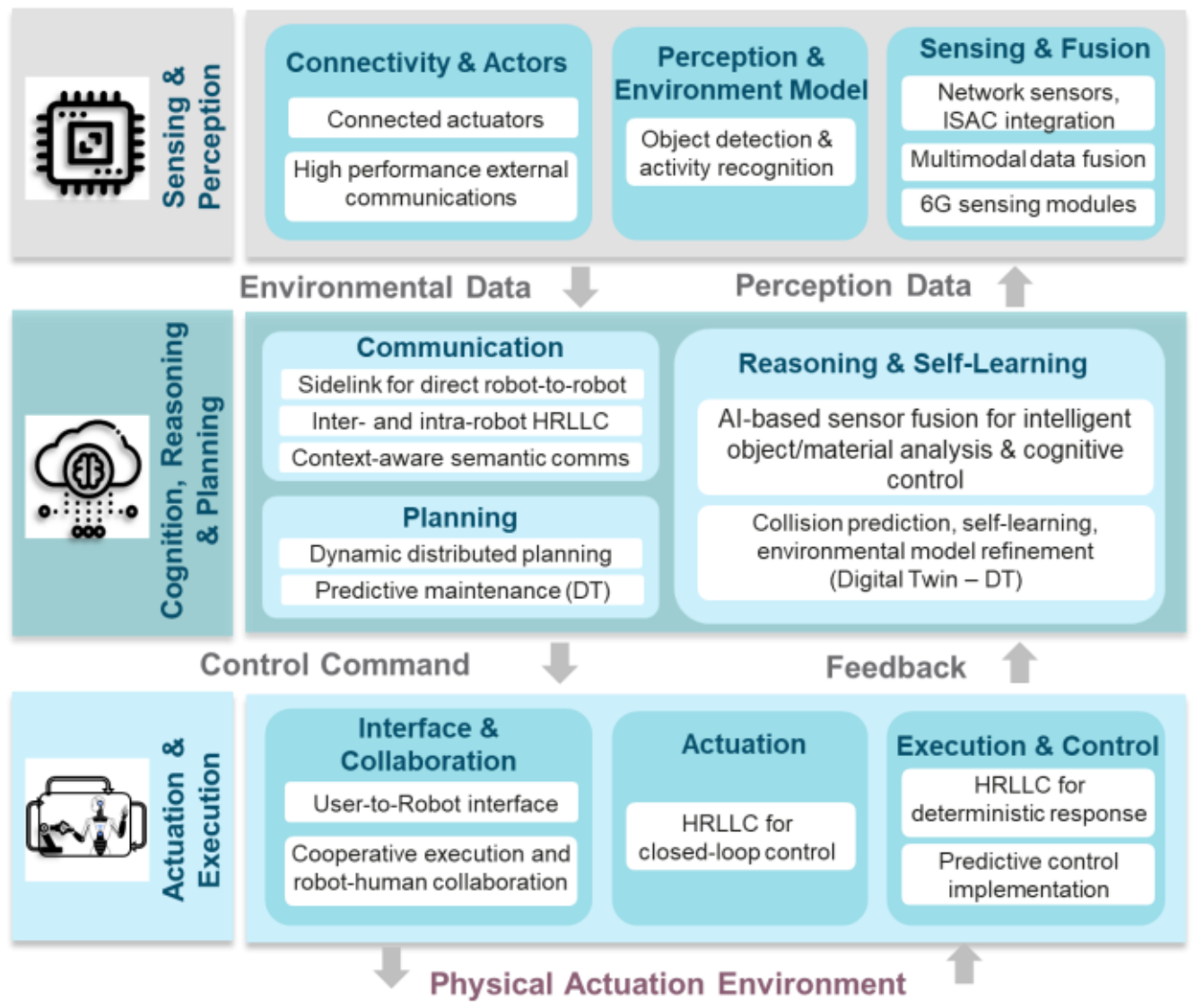

***Figure 3****. DSZ System Components and Functional Interactions*

## VI. Challenges, Conclusions, and Future Outlook

The transition from 5G to 6G entails a fundamental rethinking of network and robotic system design. 6G is not an incremental step but a transformative shift enabling more intelligent, efficient, and responsive robots. Achieving this vision requires overcoming complex technical and non-technical challenges. The integration of 6G and robotics must consider its ethical, legal, and social dimensions. Future research must integrate ethical/legal/social-by-design principles alongside technical innovation to build trust and societal acceptance.

Technical interoperability is another key challenge. Robots must operate across heterogeneous networks with ultra-low latency, reliable connectivity, and scalability for massive deployments while maintaining predictable performance. An early mapping of standards landscapes was provided in [9] and represents an important step toward systematically identifying overlaps, inconsistencies, fragmentation points, and missing specifications required for safe cross-domain integration. Such mapping supports aligning heterogeneous standards from telecommunications, robotics, AI governance, industrial automation, cybersecurity, and safety engineering domains, while clarifying interoperability requirements across the full robotic lifecycle. A dedicated data governance plan can holistically address interoperability, covering technical alignment, trustworthy data exchange, AI lifecycle governance, and proportionate regulatory compliance. A close collaboration among researchers, industry, standards bodies, and policymakers is required to develop interoperable architectures, APIs, semantic interfaces, and certification frameworks that support trustworthy 6G-enabled robotic ecosystems.

6G will reshape robotics by advancing perception, cognition, actuation, and self-learning, enabling safer and more efficient human–robot collaboration. AI-native distributed intelligence requires protocols for efficient federated learning, secure edge offloading, and intelligent resource orchestration to support scalable robotic cognition. Managing massive heterogeneous deployments will demand innovation in spectrum management, network slicing, and semantic communication to ensure scalability and efficient resource use. Researchers must also advance pervasive ISAC design by creating integrated transceiver architectures and network orchestration that jointly enable high-bandwidth communication and multimodal sensing to strengthen robotic awareness and safety while ensuring ethical and regulatory compliance. Key research challenges for the wireless communications community include refining HRLLC optimization to deliver consistent sub-millisecond reliability for safety-critical control in dynamic, interference-prone environments. Finally, developing interoperable standards (e.g., within 3GPP, IEEE, ETSI) for distributed AI, ISAC, and dynamic safety—together with collaboration with policymakers—will be essential to unlock the full potential of 6G-enabled robotics.

## Acknowledgment

The authors acknowledge the prior work [3] under the 6G Empowering Robotics Work Item within the one6G. This paper is endorsed by the one6G Association.

## References

[1] D. Kortenkamp, R. Simmons, D. Brugali, "Robotic Systems Architectures and Programming". In Siciliano, B., Khatib, O. (eds) Springer Handbook of Robotics. Springer Handbooks, 2016. doi:10.1007/978-3-319-32552-1_12

[2] R. R. Murphy, Introduction to AI robotics. MIT press, 2019.

[3] M. Ghassemian, J. Erfanian, K. Althoefer, et. al, "6G Architectural Foundations and AI-Native Solutions for Future Connected Robotics," euRobotics & One6G association whitepaper, March 2026. [Accessed on 5th May 2026] https://one6g.org/download/6873/?tmstv=1774425311

[4] W. Wu, Y. Yang, Y. Deng, A.H. Aghvami, "Goal-oriented semantic communications for robotic waypoint transmission: The value and age of information approach," IEEE Transactions on Wireless Comm., 2024.

[5] D. Gündüz, Z. Qin, I.E. Aguerri, H.S. Dhillon, Z. Yang, A. Yener, K.K. Wong, C.B. Chae, "Beyond transmitting bits: Context, semantics, and task-oriented communications," IEEE Journal on Selected Areas in Communications, 41(1), pp.5-41, 2022.

[6] Experiential Networked Intelligence (ENI); ENI System Architecture" ETSI GS ENI 005 V4.1.1 (2026-01) [Accessed on 5th May 2026] https://www.etsi.org/deliver/etsi_gs/ENI/001_099/005/04.01.01_60/gs_ENI005v040101p.pdf

[7] Recommendation ITU-R M.2160-0 (11/2023)- Framework and overall objectives of the future development of IMT for 2030 and beyond, [Accessed on 5th May 2026] https://www.itu.int/dms_pubrec/itu-r/rec/m/R-REC-M.2160-0-202311-I!!PDF-E.pdf

[8] A. M. Valenzuela, J. Silvestre-Blanes, V. M. Sempere-Paya, L. M. Bartolín-Arnau, "Multi-Access Edge Computing performance into Non-Public 5G Networks: A robot-based experiment," IEEE 20th International Conference on Factory Communication Systems (WFCS), 2024, pp. 1-8, doi: 10.1109/WFCS60972.2024.10541038.

[9] M. Ghassemian, D. Vukobratovic, *et. al.* "Standardisation landscape for 6G robotic services," IEEE CSCN, pp. 148-154, 2023.

[10] T. Zhang, G. Wang, Ch. Xue, J. Wang, M. Nixon, S. Han, "Time-Sensitive Networking (TSN) for Industrial Automation: Current Advances and Future Directions," ACM Comput. Surv. 57, 2, Article 30, 2025. doi:10.1145/3695248.

[11] Y. Li, N. Raison, S. Ourselin, T. Mahmoodi, P. Dasgupta, A. Granados, "AI solutions for overcoming delays in telesurgery and telementoring to enhance surgical practice and education," J Robot Surg. 2024 Nov 11;18(1):403. doi: 10.1007/s11701-024-02153-9.

[12] Y. Ge *et al* "Sensing with Mobile Devices through Radio SLAM: Models, Methods, Opportunities, and Challenges," IEEE Communications Magazine, vol. 63, no. 12, pp. 80-87, Dec 2025.

[13] ISO 10218- Robotics — Safety requirements — Part 1: Industrial robots [Accessed on 5th May 2026] https://www.iso.org/obp/ui/en/#iso:std:iso:10218:-1:ed-3:v1:en

[14] EU AI ACT- [Accessed on 5th May 2026] https://artificialintelligenceact.eu/

[15] ISO 13482- Robots and robotic devices — Safety requirements for personal care robots [Accessed on 5th May 2026] https://www.iso.org/standard/53820.html

## BIOGRAPHIES

**Mona Ghassemian** (Senior Member, IEEE) (mona.ghassemian@huawei.com) is principal Expert working with Huawei and a Visiting Fellow at King's College London, with over 25 years' experience shaping 6G strategy, architecture, and standards for industry verticals. She has held senior leadership roles including Senior Manager of a standardisation team at InterDigital and Principal Scientist at BT Group, alongside academic positions as Assistant and Associate Professor at KCL and University of Greenwich. She chairs the IEEE 6G empowering robotics standardisation working group.

**Andrés Meseguer Valenzuela** (IEEE Member) Andrés Meseguer Valenzuela (IEEE Member) holds a B.S. in Industrial Electronics Engineering (2018) and an M.S. in Computer Engineering and Robotics (2019) from the Universitat Politècnica de València, where he is pursuing a PhD. He is an R&D Engineer at ITI (Valencia) and has published research on mobile robotics and 5G applications in robotics.

**Ana Garcia Armada** (IEEE Fellow) (ana.garcia@uc3m.es) is a Professor at Universidad Carlos III of Madrid, Spain, where she is leading the Communications Research Group. She has been a visiting scholar at Stanford University, Bell Labs and University of Southampton. She has published more than 250 papers in international journals and conference proceedings about wireless communications and she holds seven granted patents. She has received the IEEE ComSoc/KICS Exemplary Global Service Award.

**Dejan Vukobratovic** (Senior Member, IEEE) (dejan.vukobratovic@uns.ac.rs) is a Professor at the University of Novi Sad, Serbia, and affiliated Senior Researcher at the Human-Machine Interaction group at The Institute of Artificial Intelligence Research and Development of Serbia. His research interests include wireless communication, 5G/6G networks, distributed learning, semantic communications and connected robotics.

**Periklis Chatzimisios** (Senior Member, IEEE) (pchatzimisios@ihu.gr) is a Professor at the International Hellenic University, Greece and a Research Professor at the University of New Mexico, USA, with a focus on wireless communications and networking. He is serving/having served as a Member of the IEEE 5G Initiative, IEEE ComSoc Standards Development Board, IEEE ComSoc Standards Program Development Board, IEEE ComSoc Education Services Board, IEEE ComSoc Industry Outreach Board, and is the vice-chair of the IEEE 6G empowering robotics standardisation working group.

**Kaspar Althoefer** (IEEE Fellow) (k.althoefer@qmul.ac.uk) is a Professor of robotics at Queen Mary University of London, U.K., and the Director of the Centre for Advanced Robotics. With over 30 years of experience, his research focuses on soft robotic systems, intelligent manipulation, and AI for sensor interpretation in healthcare. He was the General Chair of ICRA 2022 in London. Highly active in the 6G-robotics nexus, he led and presented the one6G White Paper on 6G Empowering Future Robotics at several events.

**Ranga Rao Venkatesha Prasad** (Senior Member, IEEE) (rvprasad@ieee.org) has been an Associate Professor with the Embedded Networked Systems Group, Delft University of Technology (TU Delft), since 2013. From 2005 to 2012, he was a Senior Researcher and an Adjunct Faculty with TU Delft. He has played a key role in bridging research and practice by contributing to standards on cognitive radios and tactile internet. His research interests include the IoT, cognitive radio, energy harvesting, sensors, communications, and robotic design. He is on the editorial board of multiple international transactions and magazines. He is the Vice-Chair of the IEEE Tactile Internet Standardisation Group.